\documentclass[a4paper,10pt]{article}
\usepackage{graphicx}
\usepackage{amssymb}
\usepackage{amsmath}
\usepackage{amscd}
\usepackage{amsthm}
\usepackage{latexsym}
\usepackage{amsfonts, array}
\usepackage{stmaryrd}
\usepackage{float}
\usepackage{subfig}
\usepackage{wrapfig}
\usepackage{color}

\title{On Computer-Intensive Simulation and Estimation Methods for Rare Event Analysis in Epidemic Models}

\author{S. Cl\'emen\c{c}on\footnote{Institut Telecom LTCI UMR Telecom ParisTech/CNRS No. 5141},\qquad 
           A. Cousien\footnote{Inserm ATIP-Avenir: "Mod\'elisation, Aide \`a la D\'ecision, et Co\^ut-Efficacit\'e en Maladies Infectieuses", Lille, France},\qquad 
           M. D\'avila Felipe\footnote{Universit\'e Pierre et Marie Curie LPMA UMR CNRS No. 7599},\qquad 
            V.C. Tran\footnote{Labo P.Painlev\'e UFR de Math\'ematiques UMR CNRS 8524, Universit\'e des Sciences et Technologies Lille 1}
}

\date{\today}

\def\N{\mathbb{N}}

\def\P{\mathbb{P}}

\def\i{\mathbf{i}}

\def\ind{{\mathchoice {\rm 1\mskip-4mu l} {\rm 1\mskip-4mu l}
{\rm 1\mskip-4.5mu l} {\rm 1\mskip-5mu l}}}

\theoremstyle{plain}
\newtheorem{theorem}{Theorem}[section]

\theoremstyle{definition}

\newtheorem{remark}[theorem]{Remark}
\newtheorem{acknowledgements}[theorem]{Remark}

\begin{document}


\maketitle

\begin{abstract}
This article focuses, in the context of epidemic models, on \textit{rare events} that may possibly correspond to crisis situations from the perspective of Public Health. In general, no close analytic form for their occurrence probabilities is available and crude Monte-Carlo procedures fail. We show how recent intensive computer simulation techniques, such as \textit{interacting branching particle methods}, can be used for estimation purposes, as well as for generating model paths that correspond to realizations of such events. Applications of these simulation-based methods to several epidemic models are also considered and discussed thoroughly.
\end{abstract}

\section{Introduction}
Since the seminal contribution of \cite{kermackmckendrick,Bartlett49}, the mathematical issues raised by the modelling and statistical analysis of the spread of communicable infectious diseases have never ceased to receive attention in the applied probability and statistics communities.  Given the great diversity of situations encountered in practice (impact of demographic phenomena, presence of control strategies, endemicity, population heterogeneity, time-varying infectivity, \textit{etc.}), a wide variety of stochastic epidemic models have been introduced in the literature, striving to incorporate more and more relevant features in order to account for real-life situations, while remaining analytically tractable. The study of the properties of the related stochastic processes (branching approximations, long-term behavior, large population asymptotics, \textit{etc.}) and the design of efficient inference methods tailored for (generally partially observed) epidemic data are still stimulating research on mathematical epidemiology. Beyond considerations of purely academic nature, many notions and techniques developed in this field are important for practitioners. Epidemic models are used to understand and control infectious diseases and their theoretical analysis sheds some light on how to come up with figures such as the reproduction number $R_0$ of the epidemics (when well-defined). From a public health guidance perspective, they can be deployed in order to simulate the likeliest scenarios or compute the probability of certain events of interest, and plan control measures to stanch a disease outbreak in real-time. However, in most situations, no close analytical form is available for these probabilities and the latter are related to events that occur very rarely, for which Crude Monte-Carlo (CMC) estimation fails.

\par It is the main purpose of this paper to review possible techniques for rare event simulation and inference in the context of epidemic models. Motivated by practical issues in Public Health, we are concerned here with critical events such as an exceedingly long duration for an epidemic, an extremely large total number of positive diagnoses (\textit{i.e.} large final size of the epidemic) in non endemic cases, the occurrence of a severe outbreak at a short horizon, \textit{etc.} Here we list a number of events that may correspond to crisis situations and express the latter as excesses of a (very large) threshold by a random variable or a (randomly stopped) stochastic process for a general class of SIR epidemic models. \textit{Importance Sampling} and \textit{Particle Filtering} methods are next adapted to tackle the problem of estimating the occurrence probabilities of these events, as well as that of simulating realizations of the latter. Beyond the description of the methodological aspects, application of these techniques for analyzing a collection of rare events related to several numerical epidemic models, some of them being fitted from real data, is also discussed.

\par The article is structured as follows. Section \ref{sec:background} introduces a general class of epidemic models, to which the simulation/estimation techniques subsequently described apply and next review events related to these models, that may correspond to health crisis situations and generally occur very rarely. Simulation-based procedures for estimating the probability of occurrence of these events are described in Section \ref{sec:methods}, while practical applications of these techniques, based on real data sets in some cases, are considered in Section \ref{sec:num} for illustration purpose. Some concluding remarks are finally collected in Section \ref{sec:concl}. In this work, it is shown that crude Monte-Carlo method often fail to provide good estimates of rare events. Importance sampling methods are a well-known alternative to estimate the occurrence probabilities of rare events. However, their efficiency relies on the choice of proper instrumental distributions, which is very complicated for most probabilistic models encountered in practice. Particle systems with genealogical selection offer an efficient computationally-based tool for estimating the targeted small probabilities.


\section{Background} \label{sec:background}
It is the goal of this section to introduce a general class of epidemic models to which the computer-intensive estimation techniques described in the subsequent section apply. The (rare) events that shall be next statistically analyzed are formulated in terms of path properties of stochastic processes.

\subsection{Epidemic models}\label{subsec:models}
The vast majority of (stochastic) epidemic models considered in the literature are of the \textit{compartmental} type. They assume that the population of interest is divided into several strata or compartments, corresponding in particular to the various possible serological statuses, and stipulate a probabilistic framework that describes the transitions from one compartment to another.

\paragraph{The Reed-Frost model.} One of the simplest epidemic models is the discrete-time chain-binomial model, generally referred to as the Reed-Frost model, that describes the spread of an infectious disease in a homogeneous and homogeneously mixing population. New infectious are assumed to occur in generations, $t=0,\; 1,\; \ldots$ and immunity is gained by the infectives of generation $t$ at generation $t+1$. Denoting by $S_t$ and $I_t$ the numbers of individuals at the $t$-th generation who are \textit{susceptible} and \textit{infective} respectively, and by $1-q$ the probability that an infective transmits the disease to a given susceptible at any generation (infections being assumed to occur independently from each other), the sequence $\{(S_t,I_t)\}_{t\in\mathbb{N}}$ with initial state $(s_0, i_0)\in \mathbb{N}^{*2}$ is a Markov chain with transitions as follows: for all $t \in \mathbb{N}$, $(s_t,i_t)$ in $\mathbb{N}^2$ and $i_{t+1}$ in $\{0,\; 1,\; \ldots,\; s_t\}$,
\begin{equation}\label{eq:trans1}
\mathbb{P}\left\{ I_{t+1}=i_{t+1} \mid (S_t, I_t)=(s_t,i_t)\right\}=\left( \begin{array}{c}s_t\\i_{t+1}
\end{array} \right) (1-q^{i_t})^{i_{t+1}}(q^{i_t})^{s_t-i_{t+1}}
\end{equation}
and
\begin{equation}\label{eq:trans2}
S_{t+1}=S_t-I_{t+1}.
\end{equation}
The set $\mathbb{N}\times\{0\}$ is \textit{absorbing} for the Markov chain $(S_t,I_t)$, meaning that the epidemics ceases as soon as the chain reaches this set (and then stays there forever), one may refer to \cite{Rev84} for an account of the Markov chain theory.
\paragraph{The standard stochastic SIR model.}
The most basic continuous-time stochastic epidemic model, generally referred to as the standard (Markovian) SIR model in a closed population of size $n$ (see the seminal contribution of \cite{Bartlett49} for instance), counts three compartments: the \textit{susceptible class} $S$, the \textit{infective class} $I$ and the \textit{removed/recovered class} $R$. This corresponds to the situation where the epidemic is of short duration, making acceptable the assumption of a closed population, and the disease provides immunity against a possible re-infection. Fig. \ref{fig:SIR} below depicts the diagram flow of this simple epidemic model (taking $\mu=\rho\equiv 0$). For clarity, we index the events $E$ through which the sizes $S(t)$, $I(t)$ and $R(t)$ of the three compartments that form the population evolve temporarily: we write $E=1$ when the event that occurs is an infection, $E=2$ when it corresponds to the removal of an infective. Taking by convention $T_0=0$ as time origin, the (continuous-time) dynamics of the model stipulates that all durations in competition are independent, infections and removals occur at time $t\geq 0$ with the rates $\lambda(S(t),I(t))=\lambda S(t)I(t)/n$ and $\gamma(I(t))=\gamma I(t)$, where $(\lambda,\gamma)\in \mathbb{R}_+^*$, respectively. Hence, the process $Z=\{(S(t),I(t),R(t))\}_{t\geq 0}$ evolves in a Markovian fashion, by jumps at random times $0<T_1<T_2<\ldots$, when events $E_1,\;E_2,\;\ldots$ in $\{1,\; 2\}$ successively occur. The dynamics can be described by stochastic differential equations driven by Poisson point measures.

\vspace{1cm}
\begin{figure}[h!]
\centering
\includegraphics[width=5cm, height=6cm,trim=5cm 7cm 5cm 7cm]{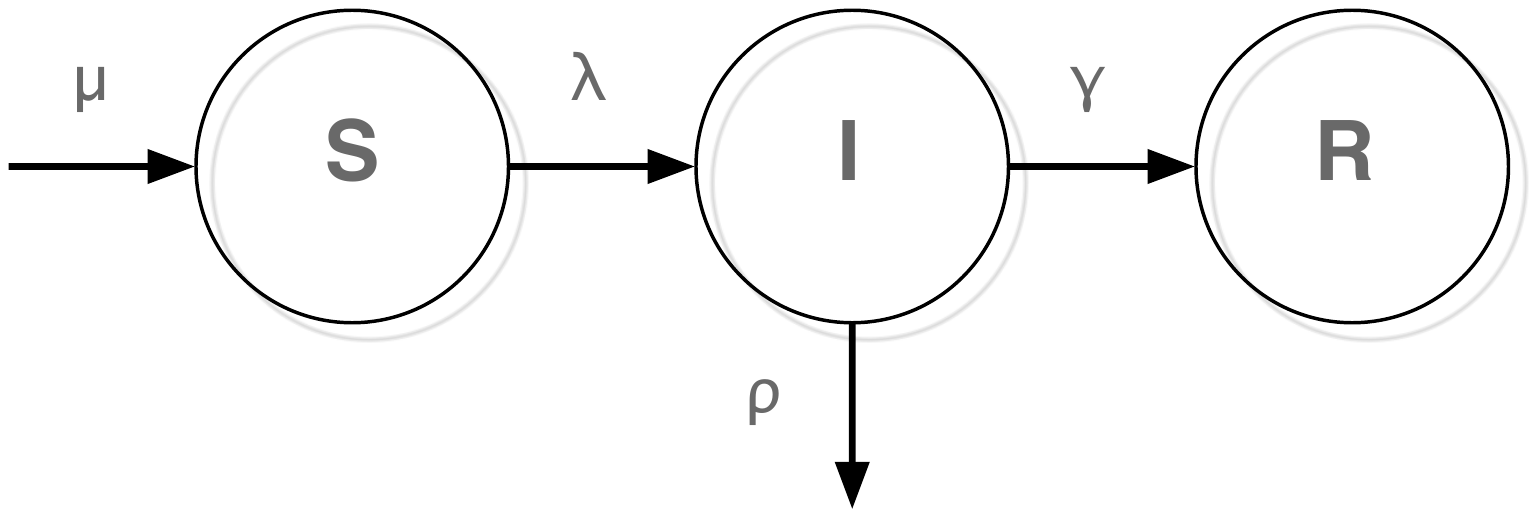}\vspace{-3cm}
\caption{Diagram flow of a basic SIR stochastic model with demography.\label{fig:SIR}}
\end{figure}
\vspace{-0.5cm}

\paragraph{Variants of the standard SIR model.}
When the epidemic under study acts on a large temporal scale, it may be necessary to incorporate additional features in the model (\textit{cf} rates $\mu$ and $\rho$ featured in Fig. \ref{fig:SIR}) accounting for the demography of the population over which the disease spreads in an endemic manner. The number and the nature of the compartments involved in the epidemic models may also vary, depending on the infectious disease considered. For instance, the SIRS model corresponds to the situation, where immunity is lost after some time, while some AIDS epidemic models count numerous compartments, in order to account for the (non exponentially distributed) AIDS incubation period (this approach is usually referred to as \textit{stage modelling}, see \cite{Isham93}). Additionally, the possible heterogeneity of the population may lead to remove the assumption of \textit{uniform mixingness} and consider instead \textit{multitype epidemic models} (refer for instance to Chapter 6 in \cite{anderssonbritton} for a review of SIR models where the population is segmented into a finite number of subcommunities) or a population \textit{structured by continuous variables} (see \cite{CDT08} for such a measure-valued stochastic process and the references therein) or spreading on random graphs which represent the underlying social network structure of the population (e.g. \cite{decreusefonddhersinmoyaltran,volz}). Indeed there are many variants of the model described above, much too numerous to be listed here exhaustively. For clarity, the problem of estimating the probability of rare events related to the spread of a transmittable disease shall be addressed in the context of simple or even simplistic models, where the epidemics is described by a discrete-time Markov chain or a jump Markov process, extensions to more general situations being straightforward in most cases.

\subsection{Rare/dramatic events in infectious disease epidemics}\label{subsec:events}

In the management of epidemics of communicable infectious diseases, the following events and quantities are of particular interest to Public Health decision makers. Here and throughout, we set $\inf \emptyset =+\infty$ by convention. The event of interest is denoted by $\mathcal{E}$. We will see that pertinent events often take the form $\mathcal{E}=\{\tau_A\leq \mathcal{T}\}$ where $A$ is a subset of the space $\mathbb{N}^3$ where the epidemics process $Z$ takes its values and where $\tau_A=\inf\{t\geq 0:\; Z(t)\in A\}$ and $\mathcal{T}$ are almost-surely finite stopping times. Hence, we are interested in level-crossing probabilities of the form:
\begin{equation}\label{target}
\mathbb{P}\left\{\tau_{A}\leq \mathcal{T}\right\}.
\end{equation}

\begin{itemize}
\item[$\bullet$] {\bf Duration of the epidemics.} In non endemic situations, the epidemics starts at a time arbitrarily set to $t=0$ and ends at a short term horizon, described by the (almost-surely finite) stopping time
$$\tau=\inf\{t\geq 0:\;\; I(t)=0\}.$$
Sharply estimating the probability $p_d(T)=\mathbb{P}\left\{ \tau > T \right\}$ that the epidemics lasts more than a (very long) period of time $[0,T]$, with $0<T<+\infty$, is an essential concern from the Public Health perspective. The computation of $1-p_d(T)$ correspond to \eqref{target} in the case where $\mathcal{T}=T$ and $A=\N\times \{0\}\times \N$.
\medskip

\item[$\bullet$] {\bf The final size of the epidemics.}
The final size of the epidemics corresponds to the total number of infected individuals between times $0$ and $\tau$ it is thus defined as the random variable $R(\tau)$. The probability $p_f(N_c)=\mathbb{P}\{R(\tau)\geq N_c\}$ that the size $R(\tau)$ exceeds a (critical) threshold value $N_c\geq 1$ smaller than $n$ in the case of a closed population of total size $n\geq 1$) is of vital interest to quantify the means to be put in place (quarantine measures, supply of medications, number of hospital beds, \textit{etc.}). Considering the stopping time $\tau_{R,N_c}=\inf\{t\geq 0:\;\; R(t)\geq N_c\}$, notice that one may write:
\begin{equation}\label{eq:rareR}
p_f(N_c)=\mathbb{P}\left\{\tau_{R,N_c}\leq \tau\right\} .
\end{equation}
$p_f(N_c)$ reduces to \eqref{target} with $\mathcal{T}=\tau$ and $A=\mathbb{N}\times \mathbb{N}\times\{N_c, \; N_c+1,\;\ldots\}$.

\medskip

\item[$\bullet$] {\bf The incidence of the epidemics. } In order to handle in real-time a crisis situation, it is relevant to consider \textit{time-dependent} quantities such that the probability that the (non cumulative) number of infectious individuals reaches a critical value $N_I$ at a certain time horizon $T<\infty$. Let $\tau_{I,N_I}=\inf\{t\geq 0:\;\; I(t)\geq N_I\}$ be the corresponding stopping time, the probability one seeks to estimate is then given by:
\begin{equation}\label{eq:rareI}
p_I(T,N_I)=\mathbb{P}\left\{  \tau_{I,N_I}\leq T\right\}.
\end{equation}
The quantity $p_I(T,N_I)$ corresponds to \eqref{target} when $\mathcal{T}=T$ and $A=\N\times \{N_I,N_I+1,\dots\}\times \N$.
\end{itemize}
Along these lines, since Public Health decision-makers often adjust their policies, depending on the number of recently diagnosed cases, one may also be interested in the following quantity, related to removed individuals (assuming by convention that, once detected, an infected individual is removed from the subpopulation of infectives): the probability that the number of cases diagnosed between times $t$ and $t+u$ increases by more than a threshold value $N_R\geq 1$, that is given by
$\mathbb{P}\left\{ R(t+u)-R(t)\geq N_R \right\}$.
 Although many other rare events of this type, related to an excessively duration or an exceeding of a large threshold, are of potential interest, given the wide variety of epidemic models (echoing the great diversity of real situations), methods for simulating rare events and estimating their probability of occurrence shall be investigated here through the examples listed above in the context of basic SIR models for the sake of simplicity.

\section{Simulation methods for rare event analysis}\label{sec:methods}

The use of Monte-Carlo simulation techniques is widespread in mathematical epidemiology, see \cite{mode} for instance. However,  crude Monte-Carlo methods (CMC) completely fail when  applied to rare events such as those listed in Section \ref{subsec:events}. We first provide in \S \ref{section:illustration} two illustrations showing the limits of CMC. An alternative in rare event simulation is known as \textit{Importance Sampling} (IS), presented in \S \ref{sec:IS}. Roughly speaking, it consists in simulating under a different probability distribution (refered to as the \textit{instrumental distribution}, equivalent to the original probability measure along a certain filtration) under which the event of interest $\mathcal{E}$ is more frequent. However, in absence of large deviation results for the vast majority of stochastic SIR models in the literature, proper instrumental distributions are difficult to obtain. In \S \ref{sec:IBS}, we present the IBS method. We describe the method and perform in Section \ref{sec:num} numeric experiments.

\subsection{Illustrations of the numerical inadequacy of CMC for simulating rare events}\label{section:illustration}
We study numerically two examples to illustrate the low quality of CMC for estimating the probabilities of rare events.

\begin{figure}[h!]
\centering
\vspace{-4cm}
\includegraphics[width=14cm, height=14cm]{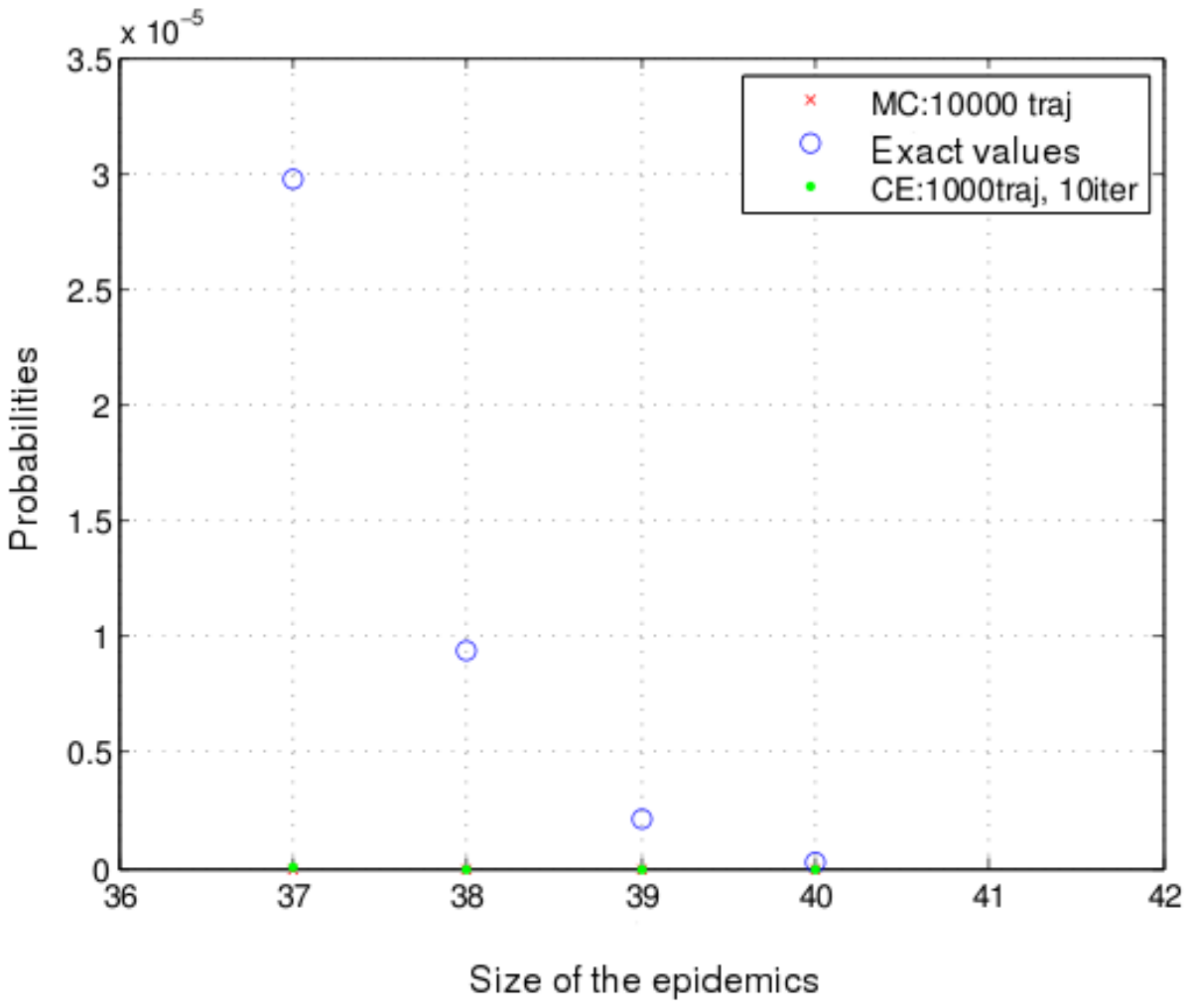}\vspace{-4cm}
\caption{In a Markovian SIR model with $(s_0,i_0)=(40,1)$ and parameters $\lambda=1$ and $\gamma=1$, crude Monte-Carlo estimate (based on $10\; 000$ replicates of the epidemics process) of the probability $p_f(N_c)$ that the size of the epidemics takes a given value are plotted as a function of $N_c$. True values are also computed.\label{fig:CMC}}
\end{figure}

First, let us consider the basic Markovian SIR model without demography (see \S \ref{subsec:models}). For this
simple model, the distribution of the final size $R(\tau)$ is proved to be the unique solution of a triangular linear system (see Theorem 2.2 in \cite{anderssonbritton} for instance, or \cite{LP90} for exact results of the same type in a more general framework), making the exact computation of the quantity $p_f(N_c)$ feasible (neglecting numerical stability issues, occurring even for moderate values of the population size $n$), whatever the threshold $N_c\geq 1$. As shown by Fig. \ref{fig:CMC}, for this particular example, the accuracy of CMC estimates of the probability $p_f(N_c)$ rapidly deteriorates when $N_c$ takes very large values (close to the total size of the population), very few (or even no) realizations of the stochastic process achieving the event $\{R(\tau)\geq N_c\}$, leading to a significant underestimation of $p_f(N_c)$, in spite of a large number of Monte-Carlo replications.
Additional comments can be found in Section \ref{sec:num}, when discussing the results.



\subsection{Importance sampling}\label{sec:IS}

A standard approach to rare event simulation is known as \textit{Importance Sampling}, see \cite{Bucklewbook} or \cite{AGbook}. The (unbiased) estimate of the probability of occurrence of the rare event is obtained by multiplying the empirical frequency of the simulations under the instrumental distribution by the likelihood ratio $\phi$, referred to as the \textit{importance function}.
For instance, when considering the standard Markovian SIR model described in the preceding section, a natural way of accelerating the occurrence of the events listed in \S \ref{subsec:events} is to speed up the infection process, while slowing down the removal (\textit{i.e.} increasing the value of the parameter $\lambda$ and decreasing that of the parameter $\gamma$). More precisely, let $\mathbb{P}$ be the probability measure under which the process $\{(S(t),I(t),R(t))\}_{t\geq 0}$ is a standard Markovian SIR model with parameters $(\lambda,\; \gamma)\in \mathbb{R}_+^{*2}$ and such that $(S(0),I(0))=(s_0,i_0)\in \mathbb{N}^{*2}$. Let $\mathbb{P}_{\text{new}}$ correspond to the pair $(\lambda_{\text{new}},\; \gamma_{\text{new}})\in \mathbb{R}_+^{*2}$, such that $\lambda_{\text{new}}\geq \lambda$ and $\gamma_{\text{new}}\leq \gamma$. Clearly, these probability measures are absolutely continuous with respect to each other along the canonical filtration $\mathcal{F}=\{\mathcal{F}_t\}_{t\geq 0}$ (\textit{i.e.} $\mathcal{F}_t$ is the $\sigma$-algebra generated by the collection of random variables $\{(S(u),I(u))\}_{u\in [0,t]}$ for all $t\geq 0$): on $\mathcal{F}_t$, the importance function (\textit{i.e.} the likelihood ratio $d\mathbb{P}/d\mathbb{P}_{\text{new}}\mid_{\mathcal{F}_t}$) is given by:
\begin{equation*}
\phi_t=\exp\left(-\int_0^t (\lambda-\lambda_{\text{new}}) S(s)I(s)/n+(\gamma-\gamma_{\text{new}}) I(s) ds\right) \left(\lambda/\lambda_{\text{new}}\right)^{N(t)-R(t)} \left(\gamma/\gamma_{\text{new}}\right)^{R(t)},
\end{equation*}
where $N(t)$ denotes the number of events $E\in\{1,\; 2\}$ occurring between times $0$ and $t$, and $T_{N(t)}$ is the last time when an event of this type occurs before time $t$. This extends to the situation where $t$ is a $\mathcal{F}$-stopping time, such as the times of exceedance considered in \S \ref{subsec:events}. Hence, if $\mathcal{E}\in \mathcal{F}_t$,  we have: $\mathbb{P}\{\mathcal{E}\}=\int \phi_t\cdot\ind\{\mathcal{E}\}d\mathbb{P}^{\text{new}}$, denoting by $\ind\{\mathcal{E}\}$ the indicator function of the event $\mathcal{E}$.

\par The success of IS crucially depends on the choice of the instrumental distribution (the specification of the instrumental parameters $(\lambda_{\text{new}},\; \gamma_{\text{new}})$ in the example previously mentioned). Ideally, it should be selected so as to reduce drastically the variance of the random variable $\phi_t\cdot \ind\{\mathcal{E}\}$, otherwise the IS approach may completely fail. Optimal choice of probability changes can be based on large-deviation techniques, when the latter are tractable for the stochastic model considered (see Chapter 5 in \cite{Bucklewbook} for further details).
However, in absence of large deviation type results for the vast majority of the stochastic SIR models considered in the literature, one faces significant difficulties for selecting importance sampling estimators with small variance in practice. Recently, a number of refinements of the IS strategy have been proposed (\textit{sequential Monte-Carlo methods} in particular), involving an iterative search of a nearly optimal instrumental distribution, see \cite{Doucetbook}. All these methods are said \textit{intrusive}, insofar as their implementation requires to call for simulation routines related to modified versions of the distribution of interest.

\noindent{\bf Cross entropy method for IS.} In the framework of estimating rare events, the \textit{cross-entropy method} (CE) introduced in \cite{Rub96} can be used to modify iteratively the instrumental distribution for estimating the occurrence probability of $\mathcal{E}$, see \cite{DBNR00,boerkroesemannorrubinstein} or \cite{ABJ06}. In the cases that are considered here, the law of the Markov processes depend on parameters: for instance $q$ in the Reed-Frost model or $(\lambda,\mu)$ in the continuous time SIR model. Let us denote by $\phi$ the set of parameters and by $\mathcal{L}(Z,\phi)$ the likelihood of the path $Z=(S_t,I_t)_{t\in \N}$ in the Reed-Frost case or $Z=(S(t),I(t),R(t))$ in the continuous time SIR model.
The idea is to choose as instrumental distribution the law $\mathcal{L}(,v)$ with the parameter $v$ that minimises the entropy with respect to the original distribution (with parameter $\phi$) conditioned on the rare event $\mathcal{E}$.
We describe the algorithm in the discrete case. The methodology also applies to the standard continuous time Markovian SIR model when it comes to estimate the quantity \eqref{eq:rareR}. Indeed, considering the embedded Markov chain $Z=(S(T_k),I(T_k))_{k\in\mathbb{N}}$, where the $T_k$'s denote the successive times when the epidemics process jumps, one may also write $p_f(N_c)=\mathbb{P}\{Z_{\tau_{\Lambda}}\in A\}$.

\par For clarity, we recall below the general principle of the CE method in the purpose of estimating the quantity $\theta=\mathbb{P}\{Z_{\tau_{\Lambda}}\in A\}$, the latter serving as a benchmark case in the experimental section, see \S \ref{subsec:toy}. Here $Z$ is a Markov chain started at $z_0$ and whose distribution is parameterized by $\phi$ and we denote by $\mathcal{L}(Z,\phi)$ its likelihood. As alternative adaptive IS methods have lead to very similar results in our experiments, they are not considered here (refer to \cite{Doucetbook}).
\medskip

\fbox{
\begin{minipage}[t]{11cm}
\medskip

\begin{center}
{\sc Adaptive Importance Sampling through the CE method}
\end{center}

\medskip
{\small

\begin{enumerate}
\item {\bf Initialization.} Set $v^{(0)}=\phi$.

\medskip

\item {\bf Iterations.} For $k=1,\;\ldots,\; K$,

\medskip
\begin{enumerate}
\item Draw $N$ sample paths starting from $x_0$ with the parameter $v^{(k-1)}$:
$$Z^{(i)}=\left(z_0,\; Z^{(i)}_1,\;\ldots,\; Z^{(i)}_{\tau^{(i)}_{\Lambda}}\right), \; \text{for }1\leq i \leq N.$$
\item Compute the IS estimate
$$
\widehat{\theta}_{k,N}=\frac{1}{N}\sum_{i=1}^N\frac{\mathcal{L}(Z^{(i)},\phi)}{\mathcal{L}(Z^{(i)},v^{(k-1)})}\cdot\ind\left\{Z^{(i)}_{\tau^{(i)}_{\Lambda}}\in A\right\},
$$
\item Define the new parameter $v^{(k)}$ as the maximum in $v$ of
$$L(v)=\frac{1}{N}\sum_{i=1}^N \ind\left\{Z^{(i)}_{\tau^{(i)}_{\Lambda}}\in A\right\}  \frac{\mathcal{L}(Z^{(i)},\phi)}{\mathcal{L}(Z^{(i)},v^{(k-1)})} \ln \mathcal{L}(Z^{(i)},v).$$

\end{enumerate}

\medskip

\item {\bf Output.}  Produce the estimate $\widehat{\theta}_{K,N}$ of the target probability.
\end{enumerate}
\bigskip
}
\end{minipage}
}

\subsection{Interacting and branching particle system methods}\label{sec:IBS}
 In contrast to the IS strategy and its variants, \textit{Interacting Branching Particle System} methods (IBPS in abbreviated form) for rare event simulation are \textit{non intrusive} in the sense that no modification of the code to run for simulating paths $Z=\{(S(t),I(t),R(t))\}_{t\geq 0}$ of the (epidemic) model under study is required. Roughly speaking, the IBPS principle as follows. We start with a population of $N$ trajectories $Z^{(1)},\;\ldots,\; Z^{(n)}$ (that we call \textit{particles}) and modify the latter in an iterative manner: paths for which the event of interest $\mathcal{E}$ "almost occurs" (in a sense that shall be specified, depending on the nature of the event $\mathcal{E}$) are ``multiplied", while the others are ``killed", following in the footsteps of the celebrated ReSTART algorithm (for Repetitive Simulated Trials After Reaching Thresholds) originally introduced in the context of teletraffic data models, see \cite{VAVA91}.

So-termed \textit{splitting techniques} (refer to \cite{GHSZ99}), thoroughly investigated in \cite{DM00} (see also \cite{CDLL06}), are fully tailored for estimating the rare event probability \eqref{target}, as well as the conditional law of the epidemics process $Z$ given the rare event of interest $\{\tau_{A}\leq \mathcal{T}\}$ is realized. The idea is to consider a sequence of increasing subsets of the state space, $A_0\supset A_1\supset A_{K+1}=A$, describing more and more difficult obstacles the process $Z$ must pass over, before reaching the target set $A$. Consider the related hitting times, defined by the recurrence relation:
$$
T_0=\inf\left\{t\geq 0:\; Z(t)\in A_0\right\}\text{ and } T_k=\inf\left\{t\geq T_{k-1}:\; Z(t)\in A_k  \right\} \text{ for }k\geq 1.
$$
We assume that $Z(0)\in A_0$ with probability one, so that $T_0=0$ almost-surely.
Clearly, the rare event probability \eqref{target} factorizes the following manner,
\begin{equation}\label{eq:FK}
\mathbb{P}\left\{T_{K+1}\leq \mathcal{T}\right\}=\mathbb{P}\left\{T_{K+1}\leq \mathcal{T}\mid T_{K}\leq \mathcal{T}\right\}\times \ldots\times \mathbb{P}\left\{T_1\leq \mathcal{T}\mid  T_{0}\leq \mathcal{T}\right\},
\end{equation}
in a product of conditional probabilities of events (hopefully) much less rare and whose realizations can be more easily simulated. The technique described subsequently precisely permits to estimate each factor in \eqref{eq:FK} and build progressively epidemics paths realizing the rare event $\{\tau_A\leq \mathcal{T}\}$ as well.

In many situations, the $A_k$'s are determined by a collection of increasing levels (the choice of the number $K$ of intermediate levels and that of the levels themselves will be discussed later, see Remark \ref{rk:adapt}). For instance, when it comes to estimate the probability $p_I(T,N_I)$ that the number of infectives exceeds a critical threshold value $N_I$ before a certain time $T<\infty$, one may consider a sequence of sublevels $0=N_I^{(0)}<\ldots<N_I^{(K+1)}=N_I$, that defines subsets $A_k=\mathbb{N}\times\{N_I^{(k)}, \; N^{(k)}_I+1,\;\ldots\}\times \mathbb{N}$ for $k=0,\;\ldots,\; K+1$. 

\begin{figure}[h!]
\centering
\vspace{-1cm}
\includegraphics[width=9cm, height=11cm]{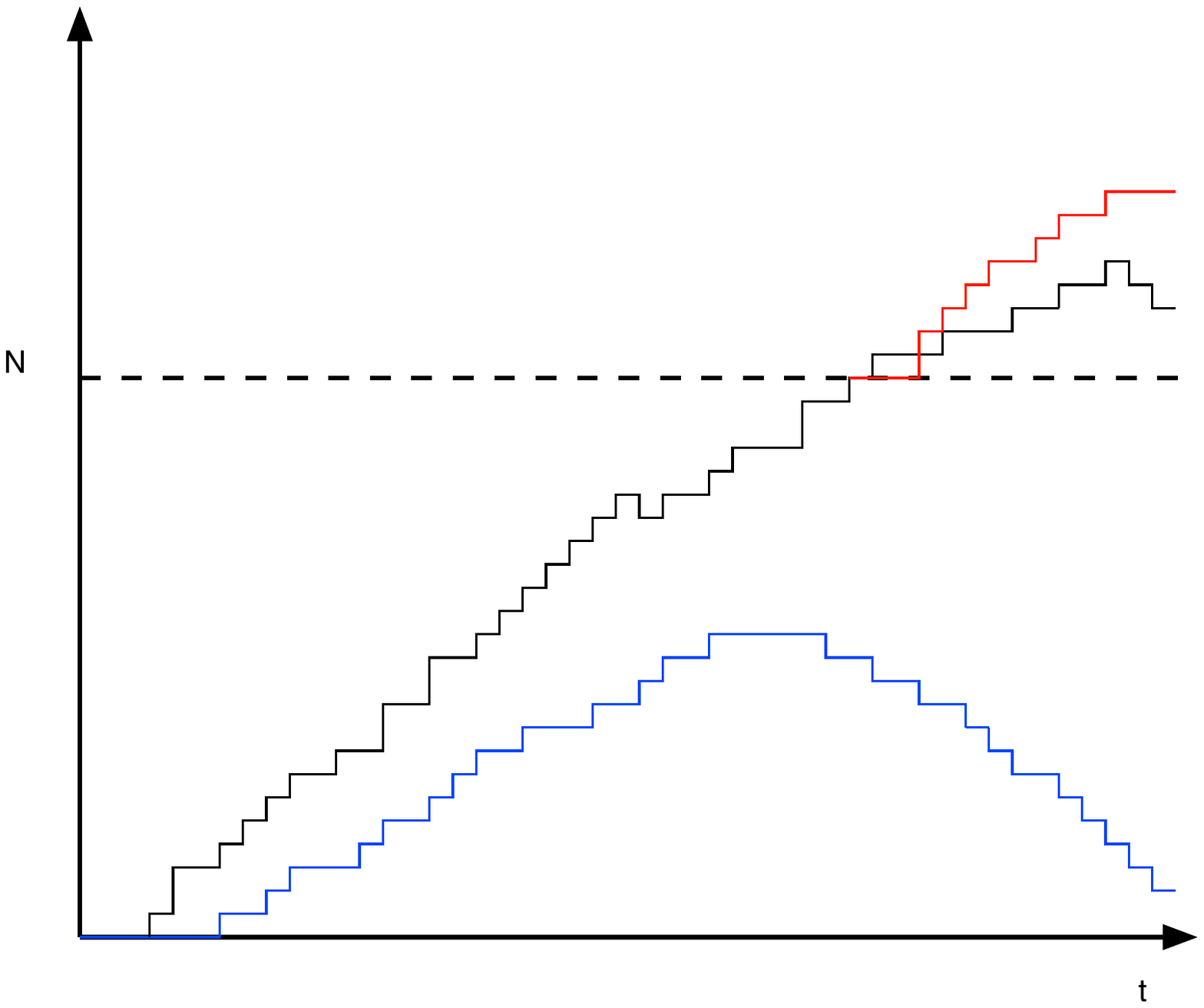}
\vspace{-4cm}
\caption{Multi-level splitting: the path in blue does not reach the current level $N$ and is thus killed, while that in black does and can be selected in order to produce an \textit{offspring}, generated by sampling from the time of exceedance (in red) \label{fig:branching}}
\end{figure}

More precisely, the particle population model evolves according to the following genealogical structure, see \cite{DM04}. At generation $k\in\{1,\;\ldots,\; K\}$,  a particle $Z$ having reached the $k$-th level before time $\mathcal{T}$ (i.e. such that $T_k\leq \mathcal{T}$) are kept while the other are deleted (\textit{selection} stage) and replaced by new particles (\textit{mutation} stage), see Fig. \ref{fig:branching}. A new particle is a novel epidemics path $Z^{\text{new}}$ whose path segment on $[0,T_k]$ coincides with that of a particle $Z$ chosen randomly among the particles such that $T_k\leq \mathcal{T}$, and whose trajectory on $[T_k,\mathcal{T}]$ (or on $[T_k,T^{\text{new}}_{k+1}]$ from a practical perspective) is simply sampled from the distribution of the epidemic process when the initial condition is $Z(T_k)$.
Of course, the algorithm stops (and is restarted) if no particle survives. Adaptive variants are described below. The \textit{selection} stage is implemented by means of \textit{weight functions} $\omega_k$ defined on the path space by $\omega_k(Z)=1$ when $T_k\leq\mathcal{T}$ and by $\omega_k(Z)=0$ otherwise. The method is then performed in $k$ steps as follows.

A quite similar approach can be considered for the estimation of the probability $p_f(N_c)$ that the total size of the epidemics rises above a large threshold $N_c\geq 1$.

\fbox{
\begin{minipage}[t]{11cm}
\medskip

\begin{center}
{\sc The IBPS algorithm}
\end{center}

\medskip
{\small

\begin{enumerate}
\item {\bf Initialization.} Start with a collection of $N\geq 1$ simulated trajectories $Z_0^{(1)},\;\ldots,\; Z_0^{(N)}$ of the epidemic process indexed by $i\in\{1,\;\ldots,\; N\}$, with the same initial condition $Z(0)=(s_0,i_0,0)$, to which the weights $\omega^{(i)}_0=1$, $1\leq i\leq N$, are assigned. Denote by $T^{(i)}_0=0<T^{(i)}_1<\ldots<T^{(i)}_{K+1}$ and $\mathcal{T}^{(i)}$ the related stopping times.

\medskip

\item {\bf Iterations.} For $k=1,\;\ldots,\; K$,

\medskip

\begin{enumerate}
\item Let $\mathcal{I}_{1,k}$ be the subset of indices $i\in \{1,\;\ldots,\; N\}$ corresponding to the epidemics paths $Z_{k-1}^{(i)}$ having reached the subset $A_k$ before time $\mathcal{T}^{(i)}$ and denote by $\#\mathcal{I}_{1,k}$ its cardinality (the algorithm is stopped and re-started if it is equal to $0$). Set $\mathcal{I}_{0,k}=\{1,\;\ldots,\; N\}\setminus \mathcal{I}_{1,k}$. For each path indexed by $i\in \mathcal{I}_{1,k}$, set $Z^{(i)}_{k}=Z^{(i)}_{k-1}$. We also define $P_k$ as the proportion of particles $Z$ that have reached the subset $A_k$ before time $\mathcal{T}$ among those which have previously reached $A_{k-1}$.

\medskip

\item For each path indexed by $i\in\mathcal{I}_{0,k}$:
\medskip

\begin{itemize}
\item ({\sc Selection step}) independently draw a particle $Z_k^{(j)}$ from distribution $\sum_{j=1}^N \omega_k^{(j)}\cdot\delta_{Z_k^{(j)}}$,
with $\omega_k^{(j)}=\omega_k(Z_k^{(j)})/(\sum_{l=1}^N\omega_k(Z_k^{(l)}))$,
\medskip

\item ({\sc Mutation step}) Define $Z^{(i)}_{k}$ as the path confounded with $Z_k^{(j)}$ until time $T^{(j)}_k$ and prolongate by simulation from the state $Z_k^{(j)}(T^{(j)}_k)$.
\end{itemize}
\item Compute $P_j=\#\mathcal{I}_{1,k} /N$ and pass onto stage $k+1$.

\end{enumerate}

\medskip

\item {\bf Output.} Compute the estimate of the target probability $\pi=\mathbb{P}\{\tau_A\leq \mathcal{T}\}$:
$$
\widehat{\pi}_N=P_1\times\ldots\times P_{K+1}.
$$
Compute also the empirical distribution
$$
\mathcal{L}_N=\frac{1}{N}\sum_{i=1}^N\delta_{Z_{K+1}^{(i)}},
$$
which may serve as an estimate of the conditional law $\mathcal{L}$ of the epidemics process given the occurrence of $\{\tau_A\leq \mathcal{T}\}$.
\end{enumerate}
\bigskip
}
\end{minipage}
}
\bigskip

\par Before showing how the IBPS performs on a variety of examples, a few remarks are in order.
\begin{remark}\label{rk:variant}{\sc (A more deterministic genetic evolution scheme)} It should be first underlined that alternative choices for the genealogical dynamics, different from that consisting in drawing uniformly among the surviving particles, could be possibly pertinent. As proposed in \cite{CDLL06} (see subsection 3.2 therein), one may also consider a $N$-particle approximation model based on the following selection/mutation scheme: in a deterministic fashion, one keeps at each stage $k$ all paths which have reached the $k$-th level, that is $N_k$ particles say. Then the other $N-N_k$ particles are killed and replaced by a particle whose path segment on $[0,T_k]$ is chosen uniformly at random among the $N_k$ "successfull" particles and completed by (independent) sampling on $[T_k,\mathcal{T}]$.
\end{remark}

\begin{remark}\label{rk:adapt}{\sc (Tuning parameters)}
Accuracy (consistency and asymptotic normality in particular) of the estimator $\widehat{\pi}_N$ produced by the IBPS algorithm has been established as the number of particles $N$ increases to infinity in \cite{CDLL06,CG07}. However, the practical implementation requires to pick several parameters: the number of intermediate levels and the levels themselves. As explained in \cite{La06}, they should be chosen, so that all factors in the product \eqref{eq:FK} are approximately of the same order of magnitude, and possibly in an adaptive way during the simulations. When applied to the problem of estimating $p_I(T,N_I)$ for instance, the adaptive variant of the multi-level splitting proposed in \cite{CG07} would consist, at each step, in sorting all the simulated paths $Z^{(i)}$ by decreasing order of the quantity $\sup_{t\in[0,T]}I^{(i)}(t)$ and take the $k$-th term as current intermediate level with fixed $k\in\{1,\;\ldots,\; N\}$ (hence killing at each step $N-k$ trajectories).\end{remark}

\begin{figure}[h!]
\centering
\vspace{-1cm}
\includegraphics[width=9cm, height=11cm]{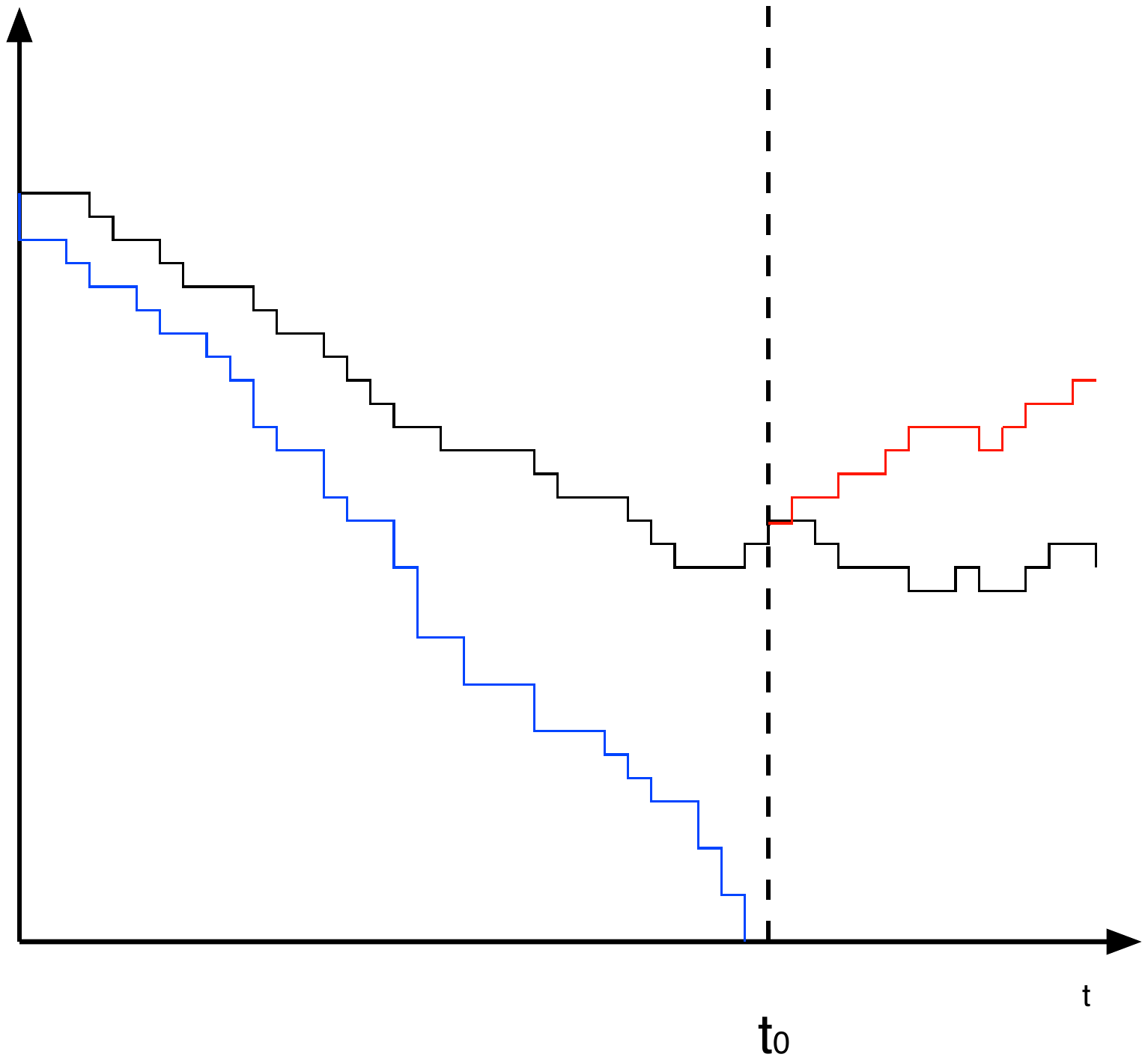}
\vspace{-4cm}
\caption{Time multi-level splitting: the path in blue extincts before time $t_0$ and is thus killed, while that in black does not and can be selected in order to produce an \textit{offspring}, generated by sampling from time $t_0$ (in red) \label{fig:time}}
\end{figure}

\begin{remark}\label{rk:time}{\sc (Persistence of the epidemics)} Observe also that the approach described above can be extended in order to estimate the probability that the epidemics lasts more than a (long) time $T>0$, $p_d(T)$. Instead of stratifying the state space of the epidemics process $Z$ (along the $I$- or $R$- axis), the idea is to write $p_D(T)=\mathbb{P}\{I(T)\geq 1\}$ and split the time axis by introducing successive durations $t_0=0<t_1<\ldots<t_{K+1}=T$ (see Fig. \ref{fig:time}). The sequence of decreasing events is then defined by $\{I(t_k)\geq1\}$ for $k=0,\;\ldots,\; K+1$ and we have:
$$
p_D(T)=\mathbb{P}\left\{I(t_{K+1})\geq 1\mid  I(t_{K})\geq 1\right\}\times \ldots\times \mathbb{P}\left\{I(t_{1})\geq 1\mid  I(t_{0})\geq 1\right\}.
$$
In this case, any particle $Z$ produces an offspring, by simulating on $[t_k,T]$ (or on $[t_k,t_{k+1}]$ in practice) a novel path segment starting from $Z(t_k)$, when it corresponds to an epidemics path that does not extinct before $t_k$, and is killed otherwise, see Fig. \ref{fig:time}. A detailed description is provided in the appendix. 
\end{remark}

\begin{remark}\label{rk:discrete}{\sc (Discrete-time models)} We point out finally that the IPBS approach can be naturally applied in a discrete-time context, so as to estimate tail probabilities $\mathbb{P}\{\sum_{k=0}^{t-1} I_k\geq N_c\}$, with $N_c\in \mathbb{N}$, at a given horizon $t\geq 1$ in a Reed-Frost model for instance. Selection/mutation steps are then performed at each intermediate time $k\in\{1,\;\ldots,\; t-1\}$: at stage $k$, $N\geq 1$ discrete paths are selected by means of a weight function $\omega_k$ defined on the path space and next mutate, through sampling of $N$ independent chains from time $k$ to time $t$. The crucial point naturally consists in a good choice for the weight functions used in the selection stage (which should be ideally based on an analysis of the variance of the corresponding estimates, when tractable). Typical choices are of the form $\omega_k(Z)=\exp(\alpha V(I_k))$ or $\omega_k(Z)=\exp(\alpha (V(I_k)-V(I_{k-1})))$, where $V:\mathbb{R}\rightarrow \mathbb{R}$ is a certain \textit{potential function} and $\alpha\geq 0$, see section \ref{sec:num} for some examples.
\end{remark}




\section{Numerical experiments}\label{sec:num}
Now that a comprehensive description of the IPBS approach has been given, it is the purpose of this section to provide strong empirical evidence that it is relevant in practice for rare event estimation in the context of (strongly Markovian) epidemics processes.

\subsection{Toy examples}\label{subsec:toy}

As a first go, we start with experiments based on simplistic epidemics models (see section \ref{sec:methods} above), in order to check the accuracy of the estimates produced by IPBS methods. For comparison purposes, CMC and (adaptive) IS estimates are also displayed. Monte-Carlo replications have been generated, so as to estimate the variability of the estimators considered as well.
\medskip

\noindent {\bf Reed-Frost model.} In this discrete-time model, we consider the probability $\P(\sum_{k=0}^{t-1} I_k>N_c)$ for $t=10$ and $N_c=90$ or $N_c=95$. Tables \ref{tb:RF1} and \ref{tb:RF2} below display estimates of this probability, together with their empirical standard deviation based on $N=1000$ Monte-Carlo replications.
The IPBS approach is here implemented with two different potential functions (\textit{cf} Remark \ref{rk:discrete}): the method referred to as $IPBS(1)$ is based on the weight function $\omega_k(Z)=\exp(\alpha V(I_k))$ with $V(I)=I$, while that referred to as $IPBS(2)$ involves $\omega_k(Z)=\exp(\alpha (V(I_k)-V(I_{k-1})))$ with $V(I)=I$. For both IPBS methods, we test $\alpha=0.1$ and $\alpha=0.01$. The levels $A_k$ appearing in the algorithms are set according to the Remark \ref{rk:adapt}: we define these levels such that at each step, a certain proportion of paths are kept (50\%, 80\% or 95\%) in our numerical example.\\

Two cases are considered, for $N_c=90$ (Table \ref{tb:RF1}) and $N_c=95$ (Table \ref{tb:RF2}). In the case $N_c=90$ the rare event has a probability estimated by CMC of 1.44e-2, while this probability is 3.0e-4 for $N_c=95$.

\begin{table}[H]
\centering
\caption{Estimates of the tail probability $\theta=\mathbb{P}\{\sum_{k=0}^{t-1} I_k\geq N_c\}$ in a Reed-Frost model, with $N_c=90$}
\begin{tabular}{| l | c c |}
\hline
Method & $\widehat{\theta}$  & s.e.\\
\hline
CMC & 1.44e-2 & (3.7e-3) \\
\hline
CE & 1.46e-2 & (1.8e-3) \\
\hline
IPBS(1) $\alpha=0.1$ 50\% & 9.1e-4 & (2.8e-4)\\
IPBS(1) $\alpha=0.01$ 50\% & 1.0e-3 & (2.6e-4)\\
IPBS(1) $\alpha=0.1$ 80\% & 1.46e-2 & (2.3e-3)\\
IPBS(1) $\alpha=0.01$ 80\% & 9.7e-3 & (1.2e-3)\\
IPBS(1) $\alpha=0.1$ 95\% & 1.42e-2 & (3.1e-3)\\
IPBS(1) $\alpha=0.01$ 95\% & 1.42e-2 & (3.1e-3)\\
\hline
IPBS(2) $\alpha=0.1$ 50\% & 1.0e-3 & (2.8e-4)\\
IPBS(2) $\alpha=0.01$ 50\% & 9.9e-4 & (2.4e-4)\\
IPBS(2) $\alpha=0.1$ 80\% & 1.0e-3 & (2.8e-4)\\
IPBS(2) $\alpha=0.01$ 80\% & 9.4e-3 & (1.7e-3)\\
IPBS(2) $\alpha=0.1$ 95\% & 1.40e-2 & (3.0e-3)\\
IPBS(2) $\alpha=0.01$ 95\% & 1.40e-2 & (3.0e-3)\\
\hline
\end{tabular}
\label{tb:RF1}
\end{table}

\begin{table}[H]
\centering
\caption{Estimates of the tail probability $\theta=\mathbb{P}\{\sum_{k=0}^{t-1} I_k\geq N_c\}$ in a Reed-Frost model, with $N_c=95$}
\begin{tabular}{| l | c c|}
\hline
Method & $\widehat{\theta}$  & s.e.\\
\hline
CMC & 3.0e-4 & (5.5e-4) \\
\hline
CE & 3.0e-4 & (1.3e-4) \\
\hline
IPBS(1) $\alpha=0.1$ 50\% & 2.0e-4 & (8.8e-5)\\
IPBS(1) $\alpha=0.01$ 50\% & 6.7e-5 & (4.2e-5)\\
IPBS(1) $\alpha=0.1$ 80\% & 4.1e-4 & (3.4e-4)\\
IPBS(1) $\alpha=0.01$ 80\% & 2.2e-4 & (2.4e-4)\\
IPBS(1) $\alpha=0.1$ 95\% & 3.2e-4 & (4.2e-4)\\
IPBS(1) $\alpha=0.01$ 95\% & 3.2e-4 & (4.2e-4)\\
\hline
IPBS(2) $\alpha=0.1$ 50\% & 1.0e-3 & (5.6e-5)\\
IPBS(2) $\alpha=0.01$ 50\% & 6.6e-5 & (4.5e-5)\\
IPBS(2) $\alpha=0.1$ 80\% & 2.5e-5 & (2.4e-4)\\
IPBS(2) $\alpha=0.01$ 80\% & 2.1e-4 & (2.3e-4)\\
IPBS(2) $\alpha=0.1$ 95\% & 3.1e-4 & (4.3e-4)\\
IPBS(2) $\alpha=0.01$ 95\% & 3.1e-4 & (4.3e-4)\\
\hline
\end{tabular}
\label{tb:RF2}
\end{table}

For both examples, we see that the estimation of CMC match with the estimation obtained by CE or by the IPBS methods when the levels are chosen such that at each step 95\% of the paths are kept. When $N_c=95$, standard deviation of the estimates are high and the obtained values are not always accurate.

\noindent {\bf Standard Markovian SIR model.} We now consider a simple continuous-time Markovian epidemics model with no demography, as described in \S \ref{subsec:models}, in the case where the target is again the tail probability related to the epidemics size, $p_f(N_c)$ namely.
We use the parameters proposed in the two examples presented in O'Neill and Roberts \cite{oneillroberts}. The first set of parameters corresponds to a toy model: $s_0=9$, $i_0=1$, $\mu\equiv 0$, $\lambda(S,I)=\lambda S I$ with $\lambda=0.12$ and $\gamma(I,R)=\gamma I$ with $\gamma=1$.
We compared the results obtained by means of the CMC, CE and IPBS methods.
Here, the method referred to as $IPBS(1)$ implements the algorithm described in the previous section, while that referred to as $IPBS(2)$ corresponds to the variant explained in Remark \ref{rk:variant}.
\begin{table}[H]
\centering
\caption{Estimates of the tail probability $\theta=p_f(N_c)$ of the size of the epidemics in a standard Markovian SIR model without demography}
\begin{tabular}{| l | c c |}
\hline
Method  & $\widehat{\theta}$ & s.e. \\
 \hline
 CMC & 2.0e-2 & (4.5e-3)\\
 \hline
 CE & 2.0e-2 & (2.5e-3) \\
 \hline
 IPBS(1) - 1\% & 2.1e-2 & (4.5e-3) \\
 IPBS(1) - 5\% & 2.1e-2 & (4.0e-3)\\
 IPBS(1) - 20\% & 2.5e-2 & (3.5e-3) \\
 \hline
 IPBS(2) - 1\% & 2.0e-2 & (4.5e-3) \\
 IPBS(2) - 5\% & 2.1e-2 & (8.0e-3) \\
 IPBS(2) - 20\% & 2.4e-2 &  (2.2e-2) \\
\hline
\end{tabular}
\label{tb:SIR1}
\end{table}

The second example in \cite{oneillroberts} comes from Bailey \cite[p.125]{bailey}. It is a smallpox outbreak in a closed community of 120 individuals in Abakaliki, Nigeria. Here the model is as above with the parameters $s_0=119$, $i_0=1$, $\lambda=0.0008254$ and $\gamma=0.087613$. The results are displayed in Table \ref{tb:SIR2}.

\begin{table}[H]
\footnotesize
\centering
\caption{Estimates of the tail probability $\theta=p_f(N_c)$ of the size of the epidemics in a standard Markovian SIR model without demography}
\begin{tabular}{| l | c c |}
\hline
Method & $\widehat{\theta}$ & s.e.\\
\hline
 CMC & 2.5e-3 & (1.6e-3)\\
 \hline
 CE & 1.6e-3 & (2.3e-4) \\
 \hline
 IPBS(1) - 1\% & 2.7e-3 & (1.3e-3)\\
 IPBS(1) - 5\% & 2.9e-3 & (9.0e-4)\\
 IPBS(1) - 20\% & 3.6e-3 & (6.7e-4) \\
 \hline
 IPBS(2) - 1\% & 2.8e-3 & (2.9e-3) \\
 IPBS(2) - 5\% & 3.1e-3 &(5.3e-3)\\
 IPBS(2) - 20\% & 3.6e-3 & (5.8e-3)\\
\hline
\end{tabular}
\label{tb:SIR2}
\end{table}

In both examples, CMC provides a good estimator of the rare probability (with 90.4\% of non-zero estimates, in the second example, \textit{i.e.} where the rare event has been observed). We take its results as a benchmark.\\

In Table \ref{tb:SIR1}, in a population of 10 individuals, we can see that every method provides a good estimate. Switching to a population of 120 individuals, one observes that CE faces difficult numerical problems related to the computation of the likelihood ratios.
This method is avoided in the sequel.\\

The IPBS method which turns out to be the more robust is the IPBS method 1, where the levels are defined so that 1\% of the paths are kept. In contrast to the Reed-Frost example, where the IPBS methods which work best correspond to a high proportion of kept trajectories (95\%), here the methods that give the results which match the best CMC correspond to those where only 1\% of the path at each iteration are kept. This may be explained by the number of iterations needed. IPBS for Reed-Frost model is implemented with a constant number of iterations, which is the number of time steps until $t$. Being too restrictive, we obtain only zero as conditional probability estimates. For the continuous time SIR model, the number of iterations is directly linked to the proportion of kept paths. The algorithm stops when the fixed proportion of best paths reaches the level $N_c$. When keep too many paths, the iteration becomes lengthy.

\subsection{An age-structured HIV epidemic model with contact-tracing}
We now consider a numerical individual-centered epidemic model, proposed and studied in the context of an asymptotically large population by \cite{CDT08}, which is effectively used for anticipating the spread of HIV in Cuba and has been statistically fitted by the means of \textit{Approximate Bayesian Computation} techniques (see \cite{BT10} for further details) based from the HIV data repository described at length in \cite{Auvert}. Experiments are naturally (and fortunately) impossible in the context of epidemics. The capacity to simulate events of interest and estimate their probability of occurrence is thus of prime importance, in order to compare the effects of different control strategies for instance. Here we investigate the impact of the contact-tracing mechanism on the probability that, by means of the IPBS method described in the previous section.

As most realistic epidemics models really used by practitioners, it is more complex than the standard Markovian SIR model with demography recalled in subsection \ref{subsec:models}, though based on the same general concepts. Precisely, this model accounts for the effect of the contact-tracing detection system set-up since 1986 in order to control the HIV epidemics accross the island by stipulating a \textit{structure by age} on the class $R$ (corresponding to the individuals diagnosed as HIV positive). The $R$ subpopulation is hence described by a \textit{point measure} $R_t$ indicating the time points since each individual in the $R$ compartment has been identified by the public health system as infected, \textit{i.e.} $R_t([a_1,a_2])$ represents the number of positive diagnoses between times $t-a_2$ and $t-a_1$ for all $0\leq a_1<a_2<+\infty$.
Apart from this, the (Markovian) dynamics of the epidemics process $\{(S(t),I(t), R_t(da))\}$ is described by the flow diagram in Fig. \ref{fig:SIR} with $\mu\equiv 0$, $\lambda(S,I)=\lambda SI$ and $\gamma(I,R)=\gamma_1I+\gamma_2 I\int_{a=0}^{+\infty} \exp(-c a) R(da)$ with $\lambda=5.4\; 10^{-8}$, $\rho\equiv 0\; 10^{-6}$, $\gamma_1=0.13$, $\gamma_3=0.19$ and $c=1$. The second term involved in the rate $\gamma(I,R)$ models the way detected individuals contribute to contact-tracing detection (notice incidentally that the smaller the parameter $c$, the more difficult the early stages of search for contact, refer to \S 2.1 in \cite{CDT08}).
\medskip

Our purpose is to estimate $p_f(N_c)$ for various values of $N_c$: $8500$, $8800$ and $9000$. As previously, IPBS is obtained with 1000 particles. For the CMC, 10e6 simulations have been performed. This permits to obtain a good estimate of the small probability $p_f(N_c)$ but also to compare CMC to IPBS. Indeed, if we separate the 10e6 simulations into 1000 runs of 1000 simulations, this allows us to count how many times the run provides an estimate equal to zero (the rare event has not been observed). As shown in Table \ref{tb:ct}, the CMC fails for the two last cases: whereas for $N_c=8500$, only 2.4\% of the simulations lead to an empirical probability equal to 0, this proportion is 84.4\% and 98.6\% for $N_c=8800$ and $N_c=9000$. This emphasizes the importance of the IPBS methods. CE methods do not give good results on such large populations, the computation of likelihood ratios being very sensitive numerically.

\begin{table}[H]
\centering
\caption{Estimates of the tail probability $\theta=p_f(N_c)$ of the size of the age-structured epidemics model with contact-tracing for Cuban HIV epidemic}
\begin{tabular}{| l c c |}
\hline
 Method & $\widehat{\theta}$ & (s.e.)\\
\hline
\multicolumn{3}{|c|}{$N_c=8500$}\\
\hline
CMC & 3.4e-3 & (1.8e-3)\\
IPBS1 - 1\% & 3.5e-3 & (1.7e-3)\\
IPBS2 - 1\% & 3.5e-3 & (3.8e-3)\\
\hline
\multicolumn{3}{|c|}{$N_c=8800$}\\
\hline
CMC & 1.7e-4 & (4.0e-4)\\
IPBS1 - 1\% & 1.5e-4 & (3.0e-4)\\
IPBS2 - 1\% & 1.7e-4 & (9.7e-4)\\
\hline
\multicolumn{3}{|c|}{$N_c=9000$}\\
\hline
CMC & 1.4e-5 & (1.2e-4)\\
IPBS1 - 1\% & 4.3e-6 & (4.4e-5)\\
IPBS2 - 1\% & 8.4e-6 & (2.1e-4)\\
\hline
\end{tabular}
\label{tb:ct}
\end{table}

\section{Conclusion}\label{sec:concl}

Though (fortunately) rare, crisis situations related to the spread of a communicable infectious disease, are of great concern to public-health managers. However, proper use of simulation-based statistical methods tailored for the estimation of such rare events is not well-documented in the mathematical epidemiology literature. Indeed, the vast majority of analyses focus on the likeliest scenarios, on events occurring with large or even overwhelming probability (\textit{e.g.} a large outbreak when the basic reproduction number is larger than one). In contrast, the present article provides an overview of recent techniques for rare event probability estimation and simulation in the context epidemics models and show how they can be used practically in order to provide efficient risk assessment tools for public-health management. The numerical results displayed in this paper 
provides strong empirical evidence that simulation methods based on interacting and branching particle systems are quite promising for this specific purpose.

\begin{acknowledgements}
The authors are grateful to Prof. H. de Arazoza for his helpful comments. The authors acknowledge support by the French
Agency for Research under the grant funding the research project {\sc Viroscopy} (ANR-08-SYSC-016-02). A.C. and V.C.T. have additional support by the Labex CEMPI (ANR-11-LABX-0007-01). The PhD of A.C. is supported by the Agence Nationale de Recherches sur le Sida et les hépatites virales (ANRS) through the project 12376.
\end{acknowledgements}

\section*{Appendix - Temporal multilevel splitting}
Here we show that the branching particle model sketched in Remark \ref{rk:time} can be used for estimating the probability $p_d(T)$ introduced in \S \ref{subsec:events}. More generally, we consider a continuous-time strong Markov process $Z=\{Z(t)\}_{t\geq 0}$ taking its values in a measurable space $E$ with initial state $z_0\in E$ and a Harris recurrent set $B\subset E$. Let $\tau_B=\inf\{t>0:\; Z(t) \in B\}$ denote the hitting time to the set $B$. Our goal is here to estimate the tail probability $\pi=\mathbb{P}\{\tau_B> t\}$, \textit{i.e.} the probability that the hitting time $\tau_B$ exceeds the (large) threshold value $t>0$, by the means of time sublevels $t_0=0<t_1<\ldots<t_K<t_{K+1}=t$. At each stage $k$, the selection step simply consists in drawing with replacement among the paths $Z$ that have not reached $B$ before time $t_k$: we set $\omega_k(Z)=1$ in this case and $\omega_k(Z)=0$ otherwise.
\medskip

\fbox{
\begin{minipage}[t]{11cm}
\medskip

\begin{center}
{\sc Temporal multilevel splitting}
\end{center}

\medskip
{\small

\begin{enumerate}
\item {\bf Initialization.} Start with a collection of $N\geq 1$ simulated trajectories $Z_0^{(1)},\;\ldots,\; Z_0^{(N)}$ of the Markov process indexed by $i\in\{1,\;\ldots,\; N\}$, with the same initial condition $z_0$ and the same weights $\omega^{(i)}_0=1$, $1\leq i\leq N$. Denote by $\tau_B^{(i)}$ the corresponding hitting times.

\medskip

\item {\bf Iterations.} For $k=1,\;\ldots,\; K$,

\medskip

\begin{enumerate}
\item Let $\mathcal{I}_{1,k}$ be the subset of indices $i\in \{1,\;\ldots,\; N\}$ corresponding to the paths $Z_{k-1}^{(i)}$ which have not reached the subset $B$ before time $t_k$, \textit{i.e.} such that $\tau_B^{(i)}>t_k$, and denote by $\#\mathcal{I}_{1,k}$ its cardinality (when it is equal to $0$, the algorithm is stopped and re-started). Set $\mathcal{I}_{0,k}=\{1,\;\ldots,\; N\}\setminus \mathcal{I}_{1,k}$. For each path indexed by $i\in \mathcal{I}_{1,k}$, set $Z^{(i)}_{k}=Z^{(i)}_{k-1}$.
\medskip

\item For each path indexed by $i\in\mathcal{I}_{0,k}$:
\medskip

\begin{itemize}
\item ({\sc Selection step}) independently draw a particle $Z_k^{(j)}$ from distribution $\sum_{j\in \mathcal{I}_{1,k}} \omega_k^{(j)}\cdot\delta_{Z_k^{(j)}}$,
with $\omega_k^{(j)}=1/\# \mathcal{I}_{1,k}$.
\medskip

\item ({\sc Mutation step}) Define $Z^{(i)}_{k}$ as the concatenation of the path $Z_k^{(j)}$ on $[0,t_k]$ with a path simulated from the state $Z_k^{(j)}(t_k)$ for times larger than $t_k$.
\end{itemize}
\item Compute $P_j=\mathcal{I}_{1,k}\# /N$ and pass onto stage $k+1$.

\end{enumerate}

\medskip

\item {\bf Output.} Compute the estimate of the target probability $\pi=\mathbb{P}\{\tau_B>t\}$:
$$
\widehat{\pi}_N=P_1\times\ldots\times P_{K+1},
$$
where $P_{K+1}$ is defined as the proportion of particles $Z$ that have not reached the subset $B$ before time $t$ among those which had not reached $A$ before time $t_K$.\\
Compute also the empirical distribution
$$
\mathcal{L}_N=\frac{1}{N}\sum_{i=1}^N\delta_{Z_{K+1}^{(i)}},
$$
which may serve as an estimate of the conditional law $\mathcal{L}$ of the epidemics process given the event $\{\tau_B>t\}$ occurs.
\end{enumerate}
\bigskip
}
\end{minipage}
}

\bigskip

We highlight the fact that the probability $\mathbb{P}\{\tau_B>t\}$ is actually of the same form as \eqref{target}. Indeed, this corresponds to the situation of the bivariate Markov process $\{(Z(t),t)\}_{t\geq 0}$ with the (rare) set $A=\mathbb{N}^*\times [T,\;+\infty[$ and $\mathcal{T}$ as the extinction time $\tau$. Therefore, works by \cite{CG07} may be adapted to prove consistence and asymptotic normality when the number of particles $N$ tends to infinity. In particular, an adaptive variant of the temporal multilevel splitting is as follows.
\medskip

\noindent {\bf Adaptive variant.} The method described above requires to fix in advance the number of time points and the time-points themselves, whereas, ideally, they should be determined in an adaptive fashion. We start by running $N$ independent paths of the epidemics and rank them by decreasing durations $\mathcal{T}^{(i)}$, $1\leq i\leq N$. The first threshold $t_1$ can be chosen as the duration of the $k-1$-th longest epidemics, so that $k$ paths are kept and $N-k$ are killed. For each killed path, we resample from the $k$ paths that have been kept and resimulate the part of the path after $t_1$. This allows to define recursively a system of longer and longer epidemic paths.

\providecommand{\noopsort}[1]{}

\end{document}